\documentstyle[preprint,prl,aps]{revtex}
\begin{document}
\draft
\title{Products of Random Matrices for Disordered Systems}
\author{A. Crisanti}
\address{Dipartimento di Fisica, Universit\`a ``La Sapienza'',
         I-00185 Roma, Italy}
\author{G. Paladin}
\address{Dipartimento di Fisica, Universit\`a dell'Aquila,
         I-67010 Coppito, L'Aquila, Italy}
\author{M. Serva}
\address{Dipartimento di Matematica, Universit\`a dell'Aquila,
         I-67010 Coppito, L'Aquila, Italy}
\author{A. Vulpiani}
\address{Dipartimento di Fisica, Universit\`a ``La Sapienza'',
         I-00185 Roma, Italy}
\date{\today}

\maketitle

\begin{abstract}
Products of random transfer matrices are applied to low
dimensional disordered systems to evaluate numerically extensive
quantities such as entropy and overlap probability distribution.
The main advantage is the possibility to avoid
numerical differentiation. The method works for arbitrary disorder
distributions at any temperature.
\end{abstract}

\pacs{75.10.Nr, 05.50.+q, 02.50.+s}

Products of random transfer matrices are a powerful tool for the study of low
dimensional systems \cite{CPV93}. Up to now they have been mainly used to
evaluate free energies and correlation functions.
{}From a formal point of view, the other thermodynamical
quantities can be obtained by differentiating the free energy with respect to
suitable parameters.
However,  in most of the cases, one has only numerical solutions
and, expecially at low temperatures, a very high precision is required
for a precise differentiation. In practice, it is very difficult
 to estimate quantities such as the
entropy or the specific heat even in a one dimensional disordered system.

The purpose of our letter is to introduce
a new  numerical method to evaluate thermodynamical
quantities  directly from the products of random transfer matrices avoiding
numerical differentiation.
The method is motivated by the papers of Masui {\it et al.}
\cite{MSJ89,MJWS93} on the study of the number of
metastable states in one and two dimensional Ising disordered systems.
As specific example we shall discuss the overlap probability
distribution $P(q)$ for zero and finite temperature for a one dimensional
Ising chain.
The method can be extended to higher dimensional systems
 with  short range interactions.

An Ising  chain consists of $N$ Ising spins with nearest neighbour
interactions in an external field. The interactions and/or the local field
are random quenched variables. At zero temperatures the only states which
are important are the metastable states in which the spin $\sigma_i$ is
aligned with the total magnetic field $h_i$ acting on it. If we denote by
$\sigma^{\alpha}_i$ the spin configuration in the metastable state $\alpha$,
the overlap between pairs of states is
$q_{\alpha\beta}= (1/N)\, \sum_{i}\ \sigma^{\alpha}_i\,
                                            \sigma^{\beta}_i$
where $\alpha$ and $\beta$ denote a pair of metastable states, and the sum is
extended to all the $N$ spins.
In general $q_{\alpha\beta}$ depends on the realization of disorder and on
the pair of states. The number of metastable states with a given overlap can
be obtained from \cite{FPV92,CPSV93a}
\begin{equation}
 {\cal N}_{N}(\omega,\beta)=
 \sum_{\sigma^{1}}\sum_{\sigma^{2}}
       e^{N\,\omega\,q_{12}}\,
      \prod_{i}^{1,N}\,\prod_{\alpha}^{1,2}\,
      \theta\left( h^{\alpha}_i\,\sigma^{\alpha}_i\right)\,
      e^{-\beta H^{\alpha}}.
\label{eq:nob}
\end{equation}
where $H^{1}$ and $H^{2}$ are the hamiltonian of two identical replicas
of the system.
  The step $\theta$-functions ensures that only the metastable states
are counted. The Gibbs weight $\exp(-\beta H^{\alpha})$ has been introduced to
select the energy of the metastable states. For any given value of $\beta$
the sum Eq.\ (\ref{eq:nob}) will be dominated by metastable states of defined
energy. In the same way any given value of $\omega$ selects metastable states
of defined overlap. Due to the $\theta$ function in Eq.\ (\ref{eq:nob}), the
parameter $\beta$ is not necessary related to the temperature, but it can be
seen as a Lagrange multiplier.

${\cal N}_{N}(\omega,\beta)$, the number of metastable states of a chain of
length $N$ for a given value of $\beta$ and $\omega$, can be written
as a sum over the energies and overlaps
\begin{equation}
 {\cal N}_{N}(\omega,\beta)= \sum_{q,E}\, {\cal N}_{N}(q,E)\,
                      e^{-2\,\beta\,E + N\,\omega\,q}.
\end{equation}
For large $N$ the sum is dominated by the saddle point energy $E'$ and
overlap $q'$, so that we can write
\begin{equation}
{\cal N}_{N}(\omega,\beta)\sim {\cal N}_{N}(q',E')\,
                      e^{-2\,\beta\,E' + N\,\omega\,q'}.
\label{eq:nn}
\end{equation}
Equation (\ref{eq:nob}) can be written as a product of random matrices, so
that the Oseledec theorem ensures that in the thermodynamic limit the quantity
in Eq.\ (\ref{eq:nn}) is the same for almost all realizations of disorder
\cite{Os88,CPV93}.
Consequently one has the Legendre transform
\begin{equation}
 \frac{1}{N}\,\overline{\ln {\cal N}_{N}(q,\epsilon)}=
           \frac{1}{N}\,\overline{\ln {\cal N}_{N}(\omega,\beta)} +
           2\,\beta\,\epsilon - \omega\,q
\end{equation}
where the bar denotes the average over the realizations of disorder. Here
$\epsilon\equiv\epsilon(\omega,\beta)$ is the energy per spin
and $q=q(\omega,\beta)$ the overlap selected by the chosen value of
$\omega$ and $\beta$. In particular the average overlap is obtained as
\cite{CPSV93a}
\begin{equation}
 q(\omega=0,\beta)= \overline{q}\equiv \int q\,\overline{P(q)}\, dq
\end{equation}
and is the value of $q$ where
$\lim_{N\to\infty} (1/N)\overline{\ln N_N(q,\epsilon)}$ reaches its maximum
value.

The key point is that not only
$(1/N)\overline{\ln {\cal N}_{N}(\omega,\beta)}$ but also
$q(\omega,\beta)$ and $\epsilon(\omega,\beta)$ can be computed directly from a
product of transfer matrices.

We shall denote by $\bbox{S}_i\equiv (\sigma^{1}_i, \sigma^{2}_i)$ the
pair of spins at the same site in two replicas of the system.
Equation (\ref{eq:nob}) can then be written as a chain sum over $\bbox{S}_i$,
i.e. as a product of the matrix
$\langle\bbox{S}_{i+1}|\hat{N}_{i}|\bbox{S}_{i}\rangle$. From
Eq.\ (\ref{eq:nob}) it is easy to see that $\hat{N}_{i}$
obeys the recursion relation
\begin{eqnarray}
 \langle\bbox{S}_{i+1}|\hat{N}_i|\bbox{S}_i\rangle & = &
    \sum_{\bbox{S}_{i-1}}\,
     \langle\bbox{S}_{i}|\hat{N}'_{i-1}|\bbox{S}_{i-1}\rangle\,
    e^{\omega \sigma^{1}_i\sigma^{2}_i}\, \nonumber \\
  &   & \times\,
     \prod_{\alpha}^{1,2} \theta(h^{\alpha}_i \sigma^{\alpha}_i)\,
     e^{-\beta E^{\alpha}_i}
\label{eq:recn}
\end{eqnarray}
with an initial matrix $\hat{N}'_{0}$ with all elements equal
and with their sum equal to unity.
Here $E^{\alpha}_i$ is the energy of the spin $\sigma^{\alpha}_i$.
The matrix $\hat{N}'_{i-1}$ is obtained by rescaling
$\hat{N}_{i-1}$ by the factor $n_{i-1}$ given by the sum of all its
elements. This ensures that the sum of the elements of $\hat{N}'_{i-1}$ is
one.

The matrix $\hat{N}_{i}$ gives the number of metastable states of a chain of
$i$ spins for all the possible configurations of
$\bbox{S}_{i}$ and $\bbox{S}_{i+1}$. Thus the logarithm of
the number of metastable states per spin, averaged over the disorder, is
\begin{equation}
 \frac{1}{N}\overline{\ln {\cal N}_{N}(\omega,\beta)}=
         \lim_{N\to\infty} \frac{1}{N}\,\sum_{i=1}^{N} \ln n_i
\end{equation}

The value of $q$ and $\epsilon$ is obtained from the derivative of
${\cal N}_N(\omega,\beta)$ with respect to $\omega$ and $\beta$,
respectively. These can be expressed in terms of the
matrices derivative of $\hat{N}_i$.
For example, defining the matrix $\hat{O}_i$ as
$\langle\bbox{S}_{i+1}|\hat{O}_i|\bbox{S}_i\rangle\equiv
    (\partial/\partial\omega)\,
     \langle\bbox{S}_{i+1}|\hat{N}_i|\bbox{S}_i\rangle$
the derivative of Eq.\ (\ref{eq:recn}) yields recursion relation
\begin{eqnarray}
 \langle\bbox{S}_{i+1}|\hat{O}_i|\bbox{S}_i\rangle & = &
    \sum_{\bbox{S}_{i-1}}\,
                 \left[
           \langle\bbox{S}_{i}|\hat{O}'_{i-1}|\bbox{S}_{i-1}\rangle
                  \right. \nonumber \\
      & & \phantom{===}
           \left. +
        \sigma^{1}\sigma^{2}
           \langle\bbox{S}_{i}|\hat{N}'_{i-1}|\bbox{S}_{i-1}\rangle
                           \right]\,
             \nonumber \\
              &   & \times\,
              e^{\omega \sigma^{1}_i\sigma^{2}_i}\,
     \prod_{\alpha=1,2} \theta(h^{\alpha}_i \sigma^{\alpha}_i)\,
     e^{-\beta E^{\alpha}_i}
 \label{eq:reco}
\end{eqnarray}
with the initial condition
\begin{equation}
 \langle\bbox{S}_2|\hat{O}_1|\bbox{S}_1\rangle=
   \sigma^{1}_1\sigma^{2}_1\,
     \langle\bbox{S}_2|\hat{N}_1|\bbox{S}_1\rangle.
\end{equation}
The matrix $\hat{O}'_{i-1}$ is related to $\hat{O}_{i-1}$ as
\begin{equation}
  \langle\bbox{S}_{i+1}|\hat{O}'_{i}|\bbox{S}_{i}\rangle=
              \left[
     \langle\bbox{S}_{i+1}|\hat{O}_{i}|\bbox{S}_{i}\rangle
     - o_i \langle\bbox{S}_{i+1}|\hat{N}'_{i}|\bbox{S}_{i}\rangle
              \right] / n_i
\label{eq:reno}
\end{equation}
where $o_i$ is the sum of the elements of $\hat{O}_i$.

{}From the renormalization Eq.\ (\ref{eq:reno}) follows that
the value of $q$ for the chosen value of $\omega$ and $\beta$ is given by the
weighted average of $o_i$ as
\begin{equation}
  q= \lim_{N\to\infty} \frac{1}{N}\,\sum_{i=1}^{N} \frac{o_i}{n_i}.
\end{equation}
A similar equation yields $\epsilon$.

The method can be generalized to finite temperatures. In this case we have to
count all the states, and not just the metastable ones. Thus
We should thus eliminate the $\theta$-functions in the above equations.
Moreover, in this case $\beta^{-1}$ is the temperature of the system.

In the absence of the $\theta$-functions, which relates the states in $i-1$,
$i$ and $i+1$, we can use a simpler algorithm \cite{CPSV93b}.
The equations are
formally the same, but with matrices replaced by vectors. Therefore, in
analogy with the zero temperature case, we introduce the vector
$\bbox{N}_i(\bbox{S}_{i+1})$ which now obeys the recursion relation
\begin{equation}
 \bbox{N}_i(\bbox{S}_{i+1})= \sum_{\bbox{S}_i}
        e^{-\beta E^{1}_i - \beta E^{2}_i
               + \omega \sigma^{1}_i \sigma^{2}_i}\,
        \bbox{N}'_{i-1}(\bbox{S}_i)
\label{eq:recnt}
\end{equation}
where the initial vector $\bbox{N}'_{0}(\bbox{S}_1)$ has all the elements
equal to one. At each iteration the vector is normalized so that
the sum of the elements of $\bbox{N}'_{i-1}(\bbox{S}_i)$ is one.
To evaluate $q$ and $\epsilon$ we
introduce the vectors of the derivatives of
$\bbox{N}_{i}(\bbox{S}_{i+1})$ with respect to $\omega$ and $\beta$,
respectively. For example, taking the derivative with
respect to $\omega$ of Eq.\ (\ref{eq:recnt}) leads to the vector
$\bbox{O}_i(\bbox{S}_{i+1})=
(\partial/\partial\omega)\bbox{N}_i(\bbox{S}_{i+1})$
obeying the recursion relation
\begin{eqnarray}
 \bbox{O}_i(\bbox{S}_{i+1}) & = &
    \sum_{\bbox{S}_{i}}\,\left[
      \bbox{O}'_{i-1}(\bbox{S}_{i}) +
         \sigma_i^{1}\sigma_i^{2} \bbox{N}'_{i-1}(\bbox{S}_{i})
                           \right]\, \nonumber \\
   &  & \times\,
      e^{\omega \sigma^{1}_i\sigma^{2}_i}\,
     \prod_{\alpha}^{1,2} e^{-\beta E^{\alpha}_i}
\end{eqnarray}
with initial condition $\bbox{O}'_0(\bbox{S}_1)= 0$.
The vector $\bbox{O}_i$ and $\bbox{O}'_i$ are related by
\begin{equation}
  \bbox{O}'_{i}(\bbox{S}_{i+1})=
              \left[
     \bbox{O}_{i}(\bbox{S}_{i+1})
       - o_i\, \bbox{N}'_{i}(\bbox{S}_{i+1})
              \right] / n_i
\end{equation}
where as above $n_i$ and $o_i$ are the sum of the elements of
$\bbox{N}_{i}(\bbox{S}_{i+1})$ and $\bbox{O}_{i}(\bbox{S}_{i+1})$,
respectively. The average energy $\epsilon$ is evaluated in a similar way.

The specific heat and susceptibility can also be computed with this method by
introducing the matrix or the vector of second derivative.

We have analyzed only Ising chains, but our arguments apply
to any formally one dimensional system and can be
easily generalized to two and three dimensional systems by considering
strips and bars, respectively. The numerical calculation becomes, however,
very heavy since the dimension of the transfer matrices and vectors grows
exponentially with the transversal dimension of the system.

Let us apply this method to the Ising chain in a random field
described by the hamiltonian
\begin{equation}
 H= -\sum_{i=1}^N J \sigma_i \sigma_{i+1} - \sum_{i=1}^{N} h_i\sigma_i
\end{equation}
where $h_i$ is a random local field. Without loosing in generality we can
take $J=1$. It is known \cite{PF78,DH83} that if $h_i$ are distributed with
probability distribution
\begin{equation}
 {\cal P}(h_i)= p\,\delta(h_i-h) + (1-p)\,\delta(h_i+h)
\end{equation}
with $0<p<1$ then the zero temperature entropy is different from zero.
This is peculiar of  distribution (16). In fact, if we replace the delta
functions by two sharp peaks of width $\Delta h$, then $S(T)\propto T$ for
$T\ll \Delta h$ \cite{PF78}. This drastic change can be clearly seen in the
behaviour of $\overline{q}= q(\omega=0,\beta=1/T)$ as a function of $T$, as
shown in Fig.\ \ref{fig:q}.
In Fig.\ \ref{fig:sh} we report the spin glass susceptibility
$\chi_{\rm SG}$, defined as the
derivative of $q(\omega,\beta)$ with respect to $\omega$ for $\omega=0$,
as a function of temperature for two different values of $\Delta h$. The
value of $\chi_{\rm SG}$ has been obtained by using the vector of the second
derivative of $\hat{N}_i$ with respect to $\omega$.
Finally Fig.\ \ref{fig:sq} shows the typical form of
$\lim_{N\to\infty} (1/N)\overline{\ln N_N(q,\epsilon)}$. The maximum is
attained for $q= \overline{q}$. In the figures $J=1$ and $h= 2$. The
convergence of the method is very fast.
Reasonable good numbers are obtained already with $O(10^4)$ iterations.
For $T=0$, the results are in very good agreement with the theoretical
prediction of Ref.\ \onlinecite{DH83}.

The parabolic shape of
$\lim_{N\to\infty} (1/N)\overline{\ln N_N(q,\epsilon)}$ means that the
overlap probability distribution $P(q)$ is selfaveraging and equal to
a $\delta$-function on the average value $\overline{q}$. This is a
general result for one dimensional systems with short range interactions
\cite{CPSV93b}.

We conclude by noting that since for $T\to 0$ the probability distribution
$P(q)$ remains a $\delta$-function \cite{CPSV93b}, the limit $T\to 0$ is not
singular and the results of the zero temperature method perfectly reproduce
those of the $T\to 0$ limit of the finite temperature method.

In conclusion we have described a method to evaluate all thermodynamical
quantities in terms of products of random transfer matrices avoiding numerical
differentiation. The method can be used, in principle, also for two and three
dimensional systems.  The method is not
restricted to spin systems, but it can be used in any other problems which
involves products of transfer matrices, e.g. directed polymers.

\begin{figure}
\caption{The average overlap $\overline{q}$ as a function of temperature $T$
         for $\Delta h= 0$ (a) and $\Delta h= 0.3$ (b). In both cases $J=1$
         and $h=2$.
        }
\label{fig:q}
\end{figure}
\begin{figure}
\caption{The spin glass susceptibility $\chi_{\rm SG}$ as a function of
         temperature $T$ for $\Delta h= 0$ (a) and $\Delta h= 0.3$ (b).
         In both cases $J=1$ and $h=2$.
        }
\label{fig:sh}
\end{figure}
\begin{figure}
\caption{$\lim_{N\to\infty} (1/N)\overline{\ln N_N(q,\epsilon)}$ as a
         function of $q$ for $\Delta h=0$, $T=0$ and
         $\epsilon=\epsilon_{\rm min}$, the ground state energy.
         The maximum is obtained for $q=\overline{q}$. In figure
         $J=1$ and $h=2$.
        }
\label{fig:sq}
\end{figure}
\end{document}